\documentstyle[eqsecnum,aps]{revtex}

\begin{document}
\draft
\title{A General Type of a Coherent State with Thermal Effects}
\author{Wen-Fa Lu}
\address{Department of Applied Physics, Shanghai Jiao Tong University,
Shanghai 200030, China
   \thanks{E-mail: wenfalu@online.sh.cn} }
\date{\today}
\maketitle

\begin{abstract}
Within the framework of thermofield dynamics, we construct a thermalized
coherent thermal state, which is a general type of the coherent state with the
thermal effects and can be presumably produced experimentally. The wavefunction
and the density matrix element in the coordinate representation are calculated,
and furthermore we give the probability densities, average values and variances
of the position, momentum and particle number, which in special cases are
consistent with those in the literature. All calculations are performed in the
coordinate representation.
\end{abstract}

Coherent states \cite{1}, which have played an important role in a varieties
of fields in physics \cite{2}, possess minimal uncertainty, mimic classical
motions, and can discribe the coherent light. Hence to investigate the effect
of temperature on it is interesting and useful, especially in quantum optics
and quantum electronics \cite{3}. Early in 1963, using the P representation of
density operator, Lachs considered the mixture of one-mode thermal and
coherent radiation, deriving the probability densities, average values and
variances of the position coordinate, momentum, as well as particle number, and
indicating their time evolution \cite{4}. Since then, the coherent state with
thermal effects has been investigated with several other methods, such as the
Bargmann representation of density operator \cite{5}, density matrix \cite{6},
the density-matrix method based on information theory \cite{7}, the
characteristic function \cite{8}, and Fock representation of
thermofield-dynamical state \cite{9}. These investigations gave rise to various
definitions of the coherent state with a finite temperature effect, and they
can be divided into thermalized coherent state and coherent thermal state
\cite{10}. Although these two states are transformed into each other by a
parameter transformation \cite{10} \cite{9} (1985), they are distinct states
and have their own physical senses. If a coherent device can excite a light
field from its ground state to the coherent state and a thermalizing device
can bring a light from its ground state to a thermal state, then a coherent
thermal state at a finite temperature $T$ is the output from the coherent
device whose input is a thermal state at $T$, and a thermalized coherent state
at $T$ is the output from a thermalizing device at $T$ whose input is a
coherent state. Thus the thermalized coherent state introduces a finite
temperature effect into the coherent state after displacing the ground state,
and the coherent thermal state introduces a finite temperature effect into the
coherent state before displacing the ground state.

Since the thermalized coherent state and the coherent thermal state are
investigated, a natural generalization of these states a thermalized coherent
thermal state (see the definition Eq.(6)) is interesting and worthy of
discussing. It takes the thermalized coherent state and the coherent thermal
state as its special cases. Also, it is more practical than those coherent
states with finite temperature effects in the literature, because both the
input and output of a coherent device usually meet thermal noises and mixed
with them owing to the inevitable existence of thermal noises. Meanwhile, for a
trapped ion, both the thermal and the coherent states can have been produced
experimentally \cite{11}, and hence a suitable combination of the experimental
techniques can produce in principle the thermalized coherent thermal state
experimentally.

This paper will discuss the thermalized coherent thermal state. First we will
give the definition of this state, then calculate the density matrix element in
the coordinate representation, discuss the probability densities, average
values and variances of the position, momentum and particle number, and
finally conclude this paper.

In the fixed-time Schr\"odinger picture, for a one-dimensional oscillator
\begin{equation}
H={\frac {1}{2m}}{\bf p}^2 +{\frac {1}{2}} m\omega^2 x^2
\end{equation}
with $x$ the position coordinate, ${\bf p}=-i\hbar{\frac {d}{dx}}\equiv -i\hbar
\partial_x$, $m$ the mass, and $\omega$ the angular frequency, the position
wavefunction of a coherent state $|\alpha>$ is easily derived as \cite{12}
\begin{eqnarray}
<x|\alpha> \equiv <x|D(\alpha)|0>
          &=& ({\frac {m\omega}{\pi \hbar}})^{\frac {1}{4}}
                  exp\{-i\alpha_1 \alpha_2\} \nonumber \\  &\ \ \ & \cdot
               exp\{-{\frac {m\omega}{2\hbar}}(x -
               \sqrt{{\frac {2\hbar}{m\omega}}}\alpha_1)^2
               + i \sqrt{{\frac {2 m\omega}{\hbar}}} \alpha_2 x \} \;,
\end{eqnarray}
where, $\alpha=(\alpha_1+i\alpha_2)$ is any complex number,
the displacement operator
\begin{equation}
D(\alpha)=\exp\{i\sqrt{{\frac {2m \omega}{\hbar}}}\alpha_2 x-i
                 \sqrt{{\frac {2}{m \hbar \omega}}}\alpha_1 {\bf p}\}
\end{equation}
is a theoretical equivalent to an ideal displacement device.

In order to consider thermal effects, thermofield dynamics introduces a tilde
oscillator whose Hamiltonian $\tilde{H}$, other observables and state vectors
can be written off from the corresponding ones of the physical oscillator
according to the tilde ``conjugation'':
$\widetilde{C O}\equiv C^* \tilde{O}$ \cite{13,14}. Here, $C$ is any
coefficient appeared in expressions of quantities for the physical system, $O$
any operator, the superscript $*$ means complex conjugation, and $\tilde{O}$
represents the corresponding operator for the tilde system. Exploiting the
physical and tilde oscillators, one can have the thermal vacuum \cite{13}
\begin{equation}
|0,\beta>={\cal T}(\theta)|0,\tilde{0}>  \;,
\end{equation}
where, $|0,\tilde{0}>=|0>|\tilde{0}>$ is the product of ground states of the
physical and tilde oscillators, $\beta={\frac {1}{k_b T}}$ with $k_b$ the
Boltzmann constant, and the unitary transformation ${\cal T}(\theta)$ (called
thermal transformation) is
\begin{equation}
{\cal T}(\theta)
  =exp\{i{\frac {\theta}{\hbar}}(x\tilde{\bf p}-\tilde{x}{\bf p})\}
\end{equation}
with $$\tanh[\theta(\beta)]=e^{-\beta\hbar\omega/2}.$$ Notice that any
physical operator commutes with any tilde operator, physical operators act on
physical states only, and similarly tilde operators on tildian states only.
Consequently the thermal-vacuum average value agrees with canonical ensemble
average in statistical mechanics. When $m$ is unit, the thermal vacuum can
describe the light in a thermal state with a Bose-Einstein distribution at
the finite temperature $T$. Usually, the thermal operator ${\cal T}(\theta)$
can be regarded as the theoretical equivalent to a thermal source or a thermal
noise at the temperature $T$.

Now we are at the position to define the thermalized coherent thermal state.
Whithin the framework of thermofield dynamics, the thermalized coherent state
can be defined as ${\cal T}(\theta)D(\alpha)\tilde{D}(\alpha)|0,\tilde{0}>$,
and the coherent thermal state $D(\alpha)\tilde{D}(\alpha)|0,\beta>$ \cite{9}
(1985), with $\tilde{D}(\alpha)$ the tildian counterpart of Eq.(3) and
$\tilde(\alpha)=\alpha^*$. Naturally, we can define the thermalized coherent
thermal state
\begin{equation}
|\beta_2,t,\alpha,\beta_1,0>\equiv {\cal T}(\theta_2)D(\alpha)
        \tilde{D}(\alpha){\cal T}(\theta_1)|0,\tilde{0}> \;.
\end{equation}
This state amounts to the output from a thermalizing device at $T_2$ whose
input is a coherent thermal state at $T_1$. Because
$D(\alpha){\cal T}(\theta_3)\tilde{D}(\alpha)\not=\tilde{D}(\alpha)
{\cal T}(\theta_3)D(\alpha)$, we donot believe that the state
${\cal T}(\theta_2)D(\alpha){\cal T}(\theta_3)
        \tilde{D}(\alpha){\cal T}(\theta_1)|0,\tilde{0}>$ has any definite
physical sense. Hence the state Eq.(6) is a general type of the coherent state
with thermal effects, at least, within the framework of thermofield dynamics.
Obviously, it takes the thermalized coherent state and the coherent thermal
state as its special cases. Next we shall give the time-dependent wavefunction
of Eq.(6) in the coordinate representation.

The time-dependent form of Eq.(6) $|t,\beta_2,\alpha,\beta_1,0>$ is
$\tilde{U}(t)|\beta_2,t,\alpha,\beta_1,0>$ with the time-evolution operator
$U(t)=e^{-{\frac {i}{\hbar}}(H-\tilde{H})t}$ \cite{13,15}. Untangling the
time-evolution and thermal operators \cite{15} as well as the displacement
operator \cite{12}, and employting Eqs.(9),(10) and (42) in Ref.~\cite{15},
one can obtain the wavefunction of the thermalized coherent thermal state in
the coordinate representation as
\begin{eqnarray}
<\tilde{x},x|t,\beta_2,\alpha,\beta_1,0>
   &=&\sqrt{{\frac {m\omega}{\pi \hbar}}}
                  exp\{(\cosh(\theta_1)-\sinh(\theta_1))^2
                  [({\frac {\alpha^2}{A}}
                 +{\frac {{\alpha^*}^2}{A^*}})\cos(\omega t) -2\alpha_1^2]\}
                 \nonumber  \\  &  &  \cdot
            exp\{-{\frac {m\omega}{2\hbar}}
      [(x \cosh(\Theta)- \tilde{x}\sinh(\Theta)
      - \sqrt{{\frac {2\hbar}{m\omega}}}{\frac {\alpha}{A}}
      (\cosh(\theta_1)-\sinh(\theta_1)))^2
        \nonumber  \\  &  &  +
     (\tilde{x} \cosh(\Theta)-x \sinh(\Theta)
     - \sqrt{{\frac {2\hbar}{m\omega}}}{\frac {\alpha^*}{A^*}}
     (\cosh(\theta_1)-\sinh(\theta_1)))^2] \}
\end{eqnarray}
with $A=\cos(\omega t)+i \sin(\omega t)$. In the case of $T_1=0$, this result
is identical to Eq.(43) with $n=0$ in Ref.~\cite{15}. The wavefunction Eq.(7)
contains a full information of the thermalized coherent thermal state, and from
it one can calculate various quantities for the state Eq.(6).

First we calculate the position density matrix element $\rho_{x',x}(t)$ from
the wavefunction Eq.(7). $\rho_{x',x}(t)$ should involve the physical position
coordinate only, and hence we have
\begin{eqnarray}
\rho_{x',x}(t)&\equiv & \int^\infty_{-\infty} <\beta,t,\alpha|x',\tilde{x}>
                                  <\tilde{x},x|\alpha,t,\beta> d\tilde{x}
            \nonumber  \\  & = &
            \sqrt{{\frac {m\omega}{\pi \hbar}}}
            \sqrt{{\frac {1}{\cosh(2\Theta)}}}
            exp\{{\frac {(\cosh(\theta_2)+\sinh(\theta_2))^2}
            {2\cosh(2\Theta)}}({\frac {\alpha}{A}}-{\frac {\alpha^*}{A^*}})^2\}
                 \nonumber  \\  & \ \ \ &
     \cdot exp\{-{\frac {m\omega}{4\hbar}}{\frac {1}{\cosh(2\Theta)}}
      [x+x'- \sqrt{{\frac {2\hbar}{m\omega}}}(\cosh(\theta_2)+\sinh(\theta_2))
         ({\frac {\alpha}{A}} + {\frac {\alpha^*}{A^*}})]^2
         \nonumber \\   & \ \ \ &
         -{\frac {m\omega}{4\hbar}}\coth(2\Theta)
  [x-x'- \sqrt{{\frac {2\hbar}{m\omega}}}
  {\frac {(\cosh(\theta_2)+\sinh(\theta_2))}{\cosh(2\Theta)}}
         ({\frac {\alpha}{A}} - {\frac {\alpha^*}{A^*}})]^2
         \} \;.
\end{eqnarray}
The position probability density is complex, Hermitian and time-dependent.

Taking $x'=x$ in Eq.(8), the position probability density is written off as
\begin{eqnarray}
\rho_{x,x}(t)&=&\sqrt{{\frac {m\omega}{\pi \hbar\cosh(2\Theta)}}}
      exp\{-{\frac {m\omega}{\hbar\cosh(2\Theta)}}
      [x- \sqrt{{\frac {\hbar}{2m\omega}}}(\cosh(\theta_2)
      \nonumber \\ &\ \ \ &
      +\sinh(\theta_2))
         ({\frac {\alpha}{A}} + {\frac {\alpha^*}{A^*}})]^2
         \} \;.
\end{eqnarray}
This is a Gaussian probability density, and from it one has easily the
position average value
\begin{equation}
<x>\equiv \int^\infty_{-\infty} x \rho_{x,x}dx
   =\sqrt{{\frac {\hbar}{2 m\omega}}}(\cosh(\theta_2)+\sinh(\theta_2))
         ({\frac {\alpha}{A}} + {\frac {\alpha^*}{A^*}})
\end{equation}
and the position variance
\begin{equation}
(\Delta x)^2\equiv <x^2>-<x>^2 = {\frac {\hbar}{2 m\omega}}\cosh(2\Theta) \;.
\end{equation}
Here and after, ``$<\cdots>$'' denotes
``$<\alpha,t,\beta|\cdots|\beta,t,\alpha>$''.

Furthermore, exploiting Eq.(8), one can calculate the momentum probability
density
\begin{eqnarray}
\rho_{p,p}(t)&=&\int^\infty_{-\infty} {\frac {1}{2\pi \hbar}}
    exp\{i {\frac {p x'}{\hbar}}-i {\frac {p x}{\hbar}}\} \rho_{x',x} dxdx'
    \nonumber  \\
    & = & \sqrt{{\frac {1}{\pi m \hbar \omega\cosh(2\Theta)}}}
      exp\{-{\frac {1}{m\hbar\omega\cosh(2\Theta)}}
      [p+i \sqrt{{\frac {m\hbar\omega}{2}}}(\cosh(\theta_2)
      \nonumber \\ &\ \ \ &
      +\sinh(\theta_2))
         ({\frac {\alpha}{A}} - {\frac {\alpha^*}{A^*}})]^2
         \} \;.
\end{eqnarray}
Here, $p$ is regarded as the momentum eigenvalue. The momentum average value
and variance are
\begin{equation}
<p>=-i\sqrt{{\frac {m \hbar\omega}{2}}}(\cosh(\theta_2)+\sinh(\theta_2))
         ({\frac {\alpha}{A}} - {\frac {\alpha^*}{A^*}})
\end{equation}
and
\begin{equation}
(\Delta p)^2 \equiv <p^2>-<p>^2= {\frac {m \hbar \omega}{2}}
                    \cosh(2\Theta) \;.
\end{equation}

Finally, from $\rho_{x',x}(t)$, we calculate the probability density, average
value and variance of the particle number. The probability density of the
particle number $n$ is
\begin{equation}
\rho_{n,n}(t)=<n|\rho|n>=\int^{\infty}_{-\infty} <n|x'><x|n>\rho_{x',x}(t)
             dx'dx  \;,
\end{equation}
where the position wavefunction of the number state $|n>$ is
$$<x|n>=({\frac {m\omega}{\pi \hbar}})^{\frac {1}{4}} (2^n n!)^{-\frac {1}{2}}
       \exp\{-{\frac {m\omega}{2 \hbar}}x^2\}
       H_n(\sqrt{{\frac {m\omega}{\hbar}}}x) \;.$$
Using formulae 7.374(8) and 7.377 in Ref.~\cite{16}, one can obtain
\begin{eqnarray}
\rho_{n,n}(t)&=& {\frac {1}{\cosh^2\Theta}}(\tanh(\Theta))^{2 n}
    \exp\{-(\cosh(\theta_2)+\sinh(\theta_2))^2(1-\tanh^2(\Theta))|\alpha|^2\}
    \nonumber  \\ & \ \ \ & \cdot
 L_n[-4({\frac {\cosh(\theta_2)+\sinh(\theta_2)}{\sinh(2\Theta)}})^2|\alpha|^2]
\end{eqnarray}
where, ${\cal L}_n[\cdots]$ is the Laguerre polynomial with zero order.
The average particle number can be calculated as
\begin{equation}
<{\bf n}>=\int^\infty_{-\infty} ({\bf n}\rho_{x',x}(t))\bigl|_{x'=x} dx
   ={\frac {1}{2}}\cosh(2\Theta)+(\cosh(\theta_2)+\sinh(\theta_2))^2|\alpha|^2
    - {\frac {1}{2}}\;,
\end{equation}
where ${\bf n}$ is the particle number operator
${\bf n}=\{{\frac {\hbar}{2 m \omega}} [({\frac {m\omega}{\hbar}})^2 x^2
-\partial^2_x] - {\frac {1}{2}}\}$. Similarly, the variance of the particle
number can be obtained as
\begin{eqnarray}
(\Delta n)^2&\equiv& <{\bf n}^2>-<{\bf n}>^2=\int^\infty_{-\infty}
          [({\bf n})^2\rho_{x',x}(t)]\bigl|_{x'=x} dx -<{\bf n}>^2
          \nonumber  \\  & \ \ \ &
          = (\cosh(\theta_2)+\sinh(\theta_2))^2\cosh(2 \Theta)|\alpha|^2
       +{\frac {1}{4}}(cosh^2 (2\Theta)-1) \;.
\end{eqnarray}

In conclusion, we have constructed the thermalized coherent thermal state. For
this state, the position probability density matrix element, the probability
densities, average values and variances of the position, momentum and particle
number have been calculated directly in the coordinate representation. This
state is a general type of the coherent state with thermal effects. In special
cases, the results in the present paper are consistent with those in the
literature \cite{4,5,6,7,8,9,10}. For example, when $t=0$ and $T_1=0$,
Eqs. (9) and (12) are consistent with Eq. (27) in Ref.~\cite{9}(Abe), Eqs.(10)
and (13) are equal to Eqs. (17a) and (17c) for the case of $\tilde{\gamma}=
\alpha^*$ in Ref.~\cite{9}(1989), respectively, and Eqs.(11) and (14) are
identical with Eqs.(4.8a) and (4.8b) in Ref.~\cite{7}, respectively. For
example again, in the case of $T_2=0$, Eq. (18) here is reduced to Eq.(2.10) in
Ref.~\cite{6}(Fearn). As has been stated in the beginning of this paper, the
thermalized coherent thermal state Eq.(6) introduces the thermal effects with
two temperatures both before and after displacing the ground state. Our results
indicate that the $T_1$-temperature effect before displacing the ground state
is different from the $T_2$-temperature effect after displacing the ground
state, and also the temperature effects in the thermalized coherent thermal
state are not a simple combination of the temperature effects in the
thermalized coherent state and the coherent thermal state. Comparing the
results here with those in Ref.~\cite{4}, one can find that in the case of
$T_2=0$ the results here can give rise to Eqs.(3.4), (3.5), (3.9) and (3.10) in
Ref.~\cite{4}. This implies that the mixture of coherent and thermal radiation
descibed by a P representation is equivalent to a coherent thermal state but
not to a thermalized coherent state. However, the P representation is
independent of the order in which the coherent and thermal sources were turned
on \cite{1} \cite{9}(1985), and hence, according to the P representation,
to introduce thermal effects before displacing the ground state is equivalent
to doing so after displacing the ground state. The theoretical reason why
there are this disprepancy betweem thermofield dynamics and the P
representation was explained in Ref.~\cite{9}(1985). However, we believe that
to prepare a thermalized coherent thermal state both of light and of an ion
experimentally is necessary because it will examine the results in this paper,
and in particular, is helpful for our understanding the above discrepancy.

\acknowledgments
This project was supported jointly by the President Foundation of
Shanghai Jiao Tong University and the National Natural Science Foundation of
China with grant No. 19875034.

 

\begin{references}
\bibitem{1} R. J. Glauber, Phys. Rev. {\bf 131}, 2766 (1963).
\bibitem{2} J. R. Klauder and B-S Sleagerstam ( $ed.$ ), {\it Coherent
            States : Applications in Physics and Mathematical Physics }
            ( World Scientific Publishing Co Pte Ltd., Singapore, 1985 );
            D. H. Feng, J. R. Klauder and M. R. Strayer ( $ed.$ ),
            {\it Coherent States : Past, Present and Future }
            ( World Scientific Publishing Co Pte Ltd., Singapore, 1994 ) .
\bibitem{3} A. Vourdas, Phys. Rev. A {\bf 34}, 3466 (1986); B. Saleh, {\it
            Photoelectron Statistics} ( Springer-Verlag, Berlin, 1978 ).
\bibitem{4} G. Lachs, Phys. Rev. {\bf 138}, B1012 (1965).
\bibitem{5} R. F. Bishop and A. Vourdas, J. Phys. A {\bf 20}, 3743 (1987);
            A. Vourdas and R. F. Bishop, Phys. Rev. A {50}, 3331 (1994).
\bibitem{6} H. Fearn and M. J. Collett, J. Mod. Opt. {\bf 35}, 553 (1988);
            M. J. W. Hall, Phys. Rev. A {\bf 49}, 42 (1994).
\bibitem{7} J. Aliaga, G. Crespo and A. N. Proto,
            Phys. Rev. A {\bf 42}, 618 (1990).
\bibitem{8} G. G. Emch and G. C. Hegerfeldt, J. Math. Phys. {\bf 27},
            2731 (1986).
\bibitem{9} S. M. Barnett and P. L. Knight, J. Opt. Soc. Am. B {\bf 2},
            467 (1985); A. Mann and M. Revzen,
             Phys. Lett A {\bf 134}, 273 (1989); S. Abe and N. Suzuki,
             Phys. Rev. A {\bf 41}, 4608 (1990).
\bibitem{10}J. Oz-Vogt, A. Mann and M. Revzen,
            J. Mod. Opt. {\bf 38}, 2339 (1991).
\bibitem{11}D. M. Meekhof, C. Monroe, B. E. King, W. M. Itano and
            D. J. Wineland, Phys. Rev. Lett. {\bf 76} (1996) 1796.
\bibitem{12}M. M. Nieto, Quantum Semiclass. Optics {\bf 8}, 1061 (1996), or
            Preprint LA-UR-96-1596 .
\bibitem{13}Y. Takahashi and H. Umezawa, Collective Phenomena {\bf 2},
            55 (1975); H. Umezawa, H. Matsumoto and M. Tachiki, {\it Thermo
            Field Dynamics and Condensed states} ( North-Holland, Amsterdam,
            1982 ); A. Das, {\it Finite Temperature Field Theory}
            ( World Scientific, Singapore, 1997 ).
\bibitem{14}Wen-Fa Lu, J. Phys. A {\bf 32}, 739 (1999), or
            hep-th/9807025.
\bibitem{15}Wen-Fa Lu, J. Phys. A {\bf 32}, 5037 (1999), or
             quant-ph/9809044, 1998 .
\bibitem{16} I. S. Gradshteyn and I. M. Ryzhik, {\it Table of Integrals,
           Series, and Products} ( Academic Press New York, 1980) .
\end{references}
\end{document}